# Back-n White Neutron Source at CSNS and its Applications

(The CSNS Back-n Collaboration)
Jing-Yu Tang[1,5,6,9*] and Qi An[1,2], Jiang-Bo Bai[3], Jie Bao[4], Yu Bao[5,6], Ping Cao[1,2], Hao-Lei Chen[1,2], Qi-Ping Chen[3], Yong-Hao Chen[5,6], Zhen Chen[1,2], Zeng-Qi Cui[7], Rui-Rui Fan[1,5,6], Chang-Qing Feng[1,2], Ke-Qing Gao[5,6], Xiao-Long Gao[5,6], Min-Hao Gu[1,5], Chang-Cai Han[8], Zi-Jie Han[3], Guo-Zhu He[4], Yong-Cheng He[5,6], Yang Hong[5,6,9], Yi-Wei Hu[7], Han-Xiong Huang[4], Xi-Ru Huang[1,2], Hao-Yu Jiang[7], Wei Jiang[5,6], Zhi-Jie Jiang[1,2], Han-Tao Jing[5,6], Ling Kang[5,6], Bo Li[5,6], Chao Li[1,2], Jia-Wen Li[1,10], Qiang Li[5,6], Xiao Li[5,6], Yang Li[5,6], Jie Liu[7], Rong Liu[3], Shu-Bin Liu[1,2], Xing-Yan Liu[3], Ze Long[5,6], Guang-Yuan Luan[4], Chang-Jun Ning[5,6], Meng-Chen Niu[5,6], Bin-Bin Qi[1,2], Jie Ren[4], Zhi-Zhou Ren[3], Xi-Chao Ruan[4], Zhao-Hui Song[8], Kang Sun[5,6,9], Zhi-Jia Sun[1,5,6], Zhi-Xin Tan[5,6], Xin-Yi Tang[1,2], Bin-Bin Tian[5,6], Li-Jiao Wang[5,6,9], Peng-Cheng Wang[5,6], Zhao-Hui Wang[4], Zhong-Wei Wen[3], Xiao-Guang Wu[4], Xuan Wu[5,6], Li-Kun Xie[1,10], Xiao-Yun Yang[5,6], Yi-Wei Yang[3], Han Yi[5,6], Li Yu[5,6], Tao Yu[1,2], Yong-Ji Yu[5,6], Guo-Hui Zhang[7], Lin-Hao Zhang[5,6,9], Qi-Wei Zhang[4], Xian-Peng Zhang[8], Yu-Liang Zhang[5,6], Zhi-Yong Zhang[1,2], Lu-Ping Zhou[5,6,9], Zhi-Hao Zhou[5,6,9], Ke-Jun Zhu[1,5,9]

1 State Key Laboratory of Particle Detection and Electronics
2 Department of Modern Physics, University of Science and Technology of China, Hefei 230026, China
3 Institute of Nuclear Physics and Chemistry, China Academy of Engineering Physics, Mianyang 621900, China
4 Key Laboratory of Nuclear Data, China Institute of Atomic Energy, Beijing 102413, China
5 Institute of High Energy Physics, Chinese Academy of Sciences (CAS), Beijing 100049, China
6 Spallation Neutron Source Science Center, Dongguan 523803, China
7 State Key Laboratory of Nuclear Physics and Technology, School of Physics, Peking University, Beijing 100871, China
8 Northwest Institute of Nuclear Technology, Xi'an 710024, China
9 University of Chinese Academy of Sciences, Beijing 100049, China
10 Department of Engineering and Applied Physics, University of Science and Technology of China, Hefei 230026, China

*Corresponding author, Email: tangjy@ihep.ac.cn

**Abstract**. Back-streaming neutrons from the spallation target of the China Spallation Neutron Source (CSNS) that emit through the incoming proton channel were exploited to build a white neutron beam facility (the so-called Back-n white neutron source), which was completed in March 2018. The Back-n neutron beam is very intense, at approximately $2\times10^7$ n/cm$^2$/s at 55 m from the target, and has a nominal proton beam with a power of 100 kW in the CSNS-I phase and a kinetic energy of 1.6 GeV and a thick tungsten target in multiple slices with modest moderation from the cooling water through the slices. In addition, the excellent energy spectrum spanning from 0.5 eV to 200 MeV, and a good time resolution related to the time-of-flight measurements make it a typical white neutron source for nuclear data measurements; its overall performance is among that of the best white neutron sources in the world. Equipped with advanced spectrometers, detectors, and application utilities, the Back-n facility can serve wide applications, with a focus on neutron-induced cross-section measurements. This article presents an overview of the neutron beam characteristics, the experimental setups, and the ongoing applications at Back-n.

## 1. Introduction

The China Spallation Neutron Source (CSNS) was completed in March 2018, and it is the first spallation neutron source and the largest scientific facility ever built in China. It mainly facilitates multidisciplinary research on material characterization using neutron scattering techniques [1-3]; however, it also provides other research platforms, including applications using white neutron beams, muon beams, and proton beams. The design goal of the CSNS Phase-I project (CSNS-I) is a proton beam power of 100 kW, which was achieved in February 2020. The other key parameters of the accelerator complex are a beam energy of 1.6 GeV and a repetition rate of 25 Hz. In the future upgrading phase (CSNS-II) that is planned to start in the next few years, the proton beam power will be increased to 500 kW.

Earlier studies [4-9] show that the back-streaming neutrons along the proton beam channel from the spallation target have the properties of a typical white neutron beam. With a wide energy spectrum, very high flux, and good time structure, it is very suitable for nuclear data measurements and other applications. In China, nuclear energy is considered an important composition in future energy sources, and research on advanced nuclear energy technologies, such as accelerator-driven subcritical systems, the thorium-based molten salt reactor, and the other fourth-generation reactors, are receiving strong support. Thus, there are eminent demands for an enhanced nuclear database, in which nuclear data measurements with neutrons are fundamental. There are also very high demands on specified neutron cross-section measurements from nuclear astrophysics, fundamental nuclear physics, and other research. The CSNS Back-n white neutron source project was launched in 2013 as an application extension of the CSNS facility, and a consortium of five institutions was formed to support the design and construction of the beam line and experimental setups. The Back-n was simultaneously completed with the CSNS project in March 2018. A number of physics spectrometers dedicated to different types of nuclear data measurements were planned; however, only four of them were constructed and made available for experiments in 2018 [10]. An upgraded spectrometer based on a $4\pi$ $BaF_2$ array (Gamma Total Absorption Facility II; GTAF-II) of the old version GTAF [11] was put into operation in 2019. Now, five dedicated spectrometers are available for users, four-unit $C_6D_6$ detectors [12] and GTAF-II for neutron capture cross-section (n, γ) measurements, FIXM, which is a multi-layer fast ionization chamber, [13] for fission cross-section (n, f) measurements, a neutron total cross-section spectrometer (NTOX), which currently also uses a fission chamber as a detector for the total cross-section (n, tot) measurements, and a light-charged particle detector array (LPDA) for light-charged particle emission (n, lcp) measurements [14-15]. Some users bring their own detectors, e.g. HPGe detectors, to conduct experiments at Back-n. The experimental setups are arranged in the two experimental halls, and flight distances from the spallation target of approximately 55 m (ES#1) and 76 m (ES#2) are shown in Fig. 1.

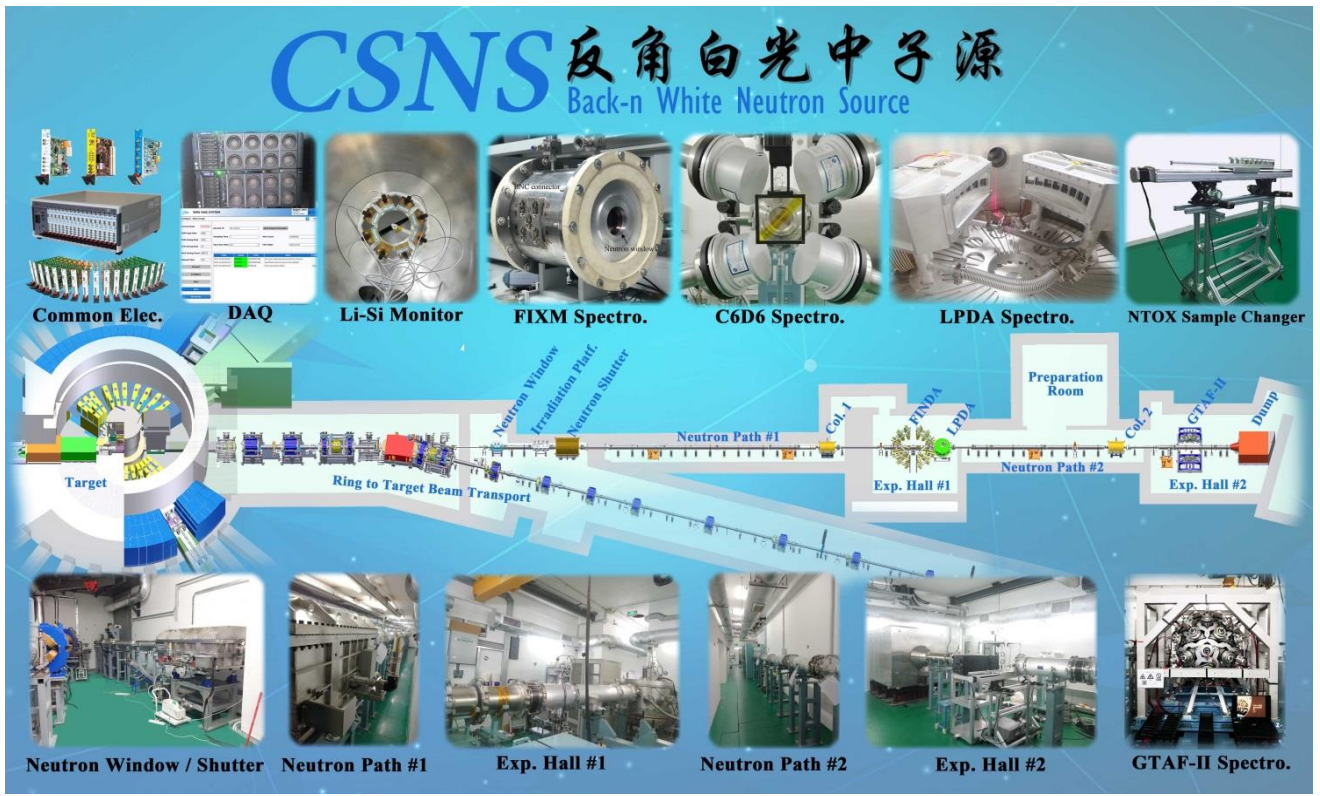

Fig. 1 Back-n white neutron beam line and spectrometers for nuclear data measurements

Table 1. Four sets of standard beam spots and neutron fluxes with relevant collimator apertures at Back-n (100 kW). The first set is primarily used to reduce fluxes, and the others are optimized according to the beam spots at ES#2. Other collimation combinations are also used frequently.

| Mode | Shutter (mm) | Coll#1 (mm) | ES#1 spot (mm) | ES#1 flux (n/cm$^2$/s) | Coll#2 (mm) | ES#2 spot (mm) | ES#2 flux (n/cm$^2$/s) |
|---|---|---|---|---|---|---|---|
| Low intensity | Φ3 | Φ15 | Φ15 | $1.3\times10^5$ | Φ40 | Φ20 | $4.6\times10^4$ |
| Small spot | Φ12 | Φ15 | Φ20 | $1.6\times10^6$ | Φ40 | Φ30 | $6.1\times10^5$ |
| Large spot | Φ50 | Φ50 | Φ50 | $1.8\times10^7$ | Φ58 | Φ60 | $6.9\times10^6$ |
| Imaging | 78×62 | 76×76 | 75×50 | $2.0\times10^7$ | 90×90 | 90×90 | $8.6\times10^6$ |

# 2 Back-n white neutron beamline

## 2.1 Beamline layout

To exploit the back-streaming neutrons, a 15° bending magnet was added at a distance of approximately 20 m from the spallation target in the proton beamline – RTBT (Ring-to-Target Beam Transport) to separate the neutrons from the proton beam. Thus, in the last RTBT section of 20 m, it is common to observe both the neutron beam and proton beam. The bending magnet is also used to remove the unwanted charged particles in the neutron beam that emit from the target and the proton beam window.

The neutron beam window close to the bending magnet starts the dedicated neutron beamline, and it separates the vacuums between the neutron beam line and RTBT. At the window, a thin foil of cadmium can be remotely driven into the beam, and it absorbs the neutrons with energies lower than 0.5 eV to avoid the overlapping between consecutive neutron pulses. Other types of absorbers with strong absorption resonances can also be used for the measurements of the in-beam neutron background. A neutron beam shutter just after the neutron window has five positions, and different apertures and can serve several functions: the neutron stopper for the radiation safety of the end-stations, the primary collimator to shape the spots at the experimental samples, and a valve to reduce neutron intensity. Together with the shutter, the two downstream collimators (Coll#1 and Coll#2) are used to shape the neutron beam profiles and adjust the intensities at the sample positions. The different sets of beam spots are shown in Table 1, and the simulations have been confirmed by the measurements [16]. In 2020, a new neutron irradiation platform between the neutron window and the neutron shutter was added to provide a high-flux ($8\times10^7$ n/cm$^2$/s) neutron irradiation service in parasitic mode, which means that the samples are arranged outside of the shutter aperture.

The shielding was carefully designed to reduce both the neutron and gamma backgrounds in the two end-station halls (ES#1, ES#2). It includes the enforced shielding wall comprising steel blocks between the RTBT tunnel and ES#1, the thick concrete walls following the shutter and the two collimators, a layer of neutron absorption material (Boron-containing polyethylene) with a thickness of 50 mm in the wall and ceil of ES#1, and a complex neutron dump in ES#2. Both the simulations [17] and measurements [18] show that a high ratio between the background and beam flux from $10^{-5}$ to $10^{-8}$ for different aperture settings can be obtained. For example, the absolute background without a sample in the beam is approximately 0.5 n/(cm$^2$.s) and 0.5 γ/(cm$^2$.s) for the off-beam background in ES#1 for a beam spot of Φ20 mm and approximately 0.01 n/(cm$^2$.s) and 0.01 γ/(cm$^2$.s) for the off-beam background in ES#2 for a beam spot of Φ30 mm. It is observed that the neutrons and gamma rays scattered by the samples are overwhelmingly dominant in the end-stations, which differ between experiments and should be solved carefully. The in-beam background from the scattered neutrons at the shutter, collimator, vacuum window, and residual gas has also been studied using both simulations and measurements.

## 2.2 Neutron beam characteristics

The characteristics of a neutron beam are extremely important for cross-section measurements and other applications. Therefore, in the design stage, the neutron production, moderation, and transport in the spallation target and the beamline have been simulated extensively using the FLUKA code [19-21] and benchmarked by the MCNP code. Comprehensive measurements were conducted in the commissioning phase [16, 22-23] and will be further pursued in the future.

*2.2.1 Neutron energy spectrum and flux*

The spallation target has 11 slices, which are made of tungsten blocks cladded with tantalum and thin cooling-water layers with a size of $160 \times 60 \times 600$ mm$^3$ (H×V× L) [24]. The three moderators above and below the target module and the beryllium reflector are also included in the simulations. The proton beam size, which is four times the root mean square size at the target entrance, is 120 (H) × 40 mm (V), and the distribution is assumed to be quasi-uniform [25]. Figure 2 shows the neutron energy spectrum at ES#2, which is approximately 76 m from the spallation target, together with the measurements. The spectrum shapes for different collimator aperture settings and flight lengths are slightly different. Two detectors were used for the measurement of the energy spectrum: a multi-layer ionization chamber with $^{235}$U samples and the Li-Si monitor, which is an array of eight silicone detectors associated with a $^6$LiF foil in-beam. The former provides consistent simulation results in the higher energy region' however, it is less accurate in the low energy region, owing to rich resonances. The latter provides good results in the lower energy range, but bad statistics in the high energy region [23]. The two detectors together provide an energy spectrum in the full-energy range. With the in-beam cadmium filter, the neutron energy spans 0.5 eV to approximately 200 MeV, with a peak at approximately 1 MeV. The absolute neutron flux was measured using an ionization chamber equipped with a mass-calibrated $^{235}$U sample, and the on-line monitoring is performed by the Li-Si monitor. The wide energy range, very high flux, and good time resolution make the neutron beam excellent for nuclear data measurements and other types of applications.

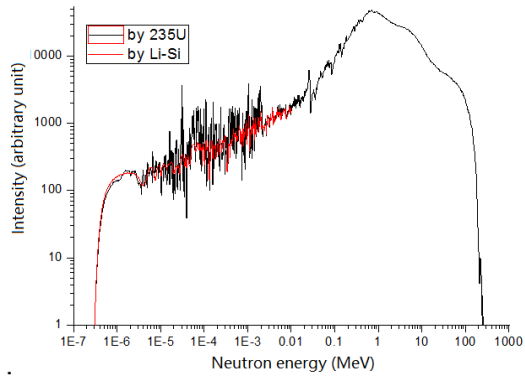

**Fig. 2.** Measured neutron energy spectra at 76 m from the target (with the Cd foil filter on), combined by an ionization chamber ($^{235}$U sample) and an Li-Si monitor ($^6$Li sample).

*2.2.2 Time-of-flight method*

The time-of-flight (TOF) method is essential to determine the neutron energy in pulsed neutron sources, including neutron scattering and white neutron applications. To obtain a good time resolution or energy resolution, a relatively short pulse length and a relatively long flight path are generally required. For some experiments, very high time resolution of a few per-mille or even lower, is required. The flight length at Back-n is limited to approximately 77 m at the far end, and a thick spallation target widens the neutron pulse width. Thus, the time resolution of a few per-mille in most of the applicable energy range is considered good, but not comparable with some beamlines with flight paths of more than 100 m at some white neutron facilities. For neutrons with energies larger than 10 keV, the time resolution is primarily determined by the proton pulse width instead of the target thickness and moderation for lower energies. One important but unique issue that is different from other white neutron beamlines in the world is that the nominal CSNS operation mode delivers a proton beam with two bunches separated by 410 ns in a pulse. This poses additional problems for the TOF measurements when neutrons with energies higher than a few keV are involved. Two measures have been taken: one is to develop the so-called unfolding method to de-convolute the two bunches in one pulse [26], and the other is to develop different operation modes for the CSNS accelerators to provide single bunch, or even short-bunch, proton beams [26]. Both the unfolding method and the single bunch operation mode have been successfully applied at Back-n [22]. The time resolutions at Back-n for different working modes are shown in Fig.3

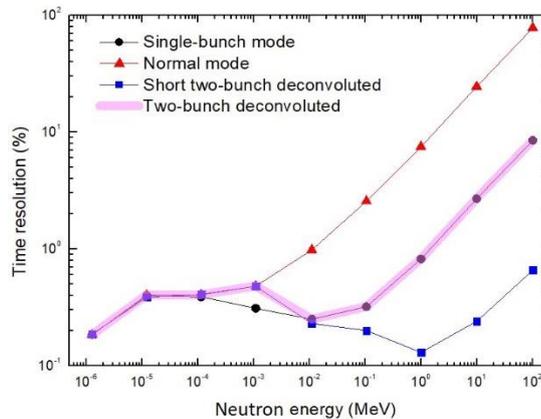

Fig. 3 Time resolutions with respect to neutron energy for different acceleration operation modes (flight path: 77 m)

The errors obtained when measuring the flight time of each detected neutron are also important. The so-called T0 signal that represents the neutron emission time from the target can be determined either from the proton beam arrival time at the target or from the measurement of the accompanying gamma burst. They reflect the T0 signal for the whole neutron pulse because it is not possible to measure a neutron without affecting it. Both methods are used at Back-n, depending on the experimental setups. Errors of a few nanoseconds can be obtained with a fast current transformer in the RTBT, and slightly worse errors are obtained with the gamma measurement. The error for the arrival signal T1 at the detector is fully dependent on the detector type, varying from less than a nanosecond to hundreds of nanoseconds.

**2.3 Accompanying gamma rays with the neutron beam**

There is a large production of gamma rays when the proton beam bombards the spallation target. They can be classified into two categories: the prompt gamma rays (the so-called gamma flash), which have the same time structure as the proton beam, and the delayed gamma rays from the activated nuclides, which have no evident time structure. The simulations and measurements show that both the yields from the prompt gamma rays and delayed gamma rays have a similar level of neutrons, e.g. approximately $10^{6-7}$ /cm$^2$/s [17-18]. The gamma rays transmitted along the back-streaming neutron beam typically pose problems to the nuclear data measurements, especially those measuring the secondary gamma rays, such as neutron capture measurements and in-beam gamma spectroscopy; however, some detectors use the gamma flash to determine the T0 for the TOF measurements. The in-beam gamma background can be largely suppressed by inserting a lead absorber with a thickness of 30–100 mm in the upstream beamline, and it reduces the neutron flux only modestly. The simulations show that the TOF measurement is only minimally affected by neutrons with energies below 1 MeV, and a detailed study, including experiments to verify this, is under way.

# 3. Experimental setups for nuclear data measurements and other applications

As mentioned above, there are two experimental halls that can be used to host different spectrometers for nuclear data measurements and other experimental set-ups, and they have distances of approximately 55 m and 77 m from the spallation target. In total, seven types of spectrometers have been planned since the beginning of the project; however, only five are available today: a 4-unit $C_6D_6$ detector and 4π $BaF_2$ detector (GTAF-II) array for neutron capture measurements, a multi-layer fission chamber (FIXM) for fission cross-section measurements, a combined detector system (NTOX) for total cross-section measurements, and a 16-unit telescope array (LPDA) for light charged particle measurements. User-owned HPGe detectors and a time projection chamber prototype can also be used for experiments. Here, the five spectrometers are described in further detail.

In addition, a time-gated CMOS camera is used for neutron imaging experiments, and one multi-functional sample support is used for detector calibration tests and irradiation applications.

*3.1 $C_6D_6$ detectors*

This spectrometer [12] is similar to the one at CERN n_TOF [28-29] (see Fig. 4). It is composed of four units of $C_6D_6$ (deuterated benzene liquid scintillators, Type: EJ315) detectors, which have high gamma detection efficiency and are less sensitive to neutrons. The sample is supported in the atmosphere, and the distance to the detector centers is approximately 150 mm. The aluminum supporting frame was designed to have the least material to reduce the influence from the scattered gamma rays and neutrons. With a low coverage solid angle, the spectrometer detects one gamma ray at a time; however, several gamma rays are typically generated in a single neutron-capture reaction. In the data treatment, it is also necessary to introduce weighting functions to compensate the detection efficiency with respect to the gamma energy. Sometimes, a correction to the detection sensitivity of scattered neutrons may be required.

*3.2 FIXM*

The FIXM [13] is an ionization chamber that has eight layers, each either posing a standard sample or samples to be measured, as shown in Fig. 5. It is similar to the ionization chamber at CERN n_TOF [30]. The standard $^{235}$U samples and $^{238}$U samples are used to calibrate the neutron flux in the whole neutron energy range, and up to six experimental samples can be used at a time. The structure design of the ionization chamber was optimized to have a fast time response, which is important in determining the neutron energy with the TOF method. The neutron energy coverage for the experiments with the FIXM is very large, from 1 eV to no less than 20 MeV. The accompanying gamma burst can also be detected in the high-pressure mode and serve as the T0 signal calibration. Despite the relatively high detection efficiency using $^{235}$U as the standard sample in the low energy range, the rich resonances of $^{235}$U induce some problems in obtaining cross-sections with high precision, which is predicted to be less than 3%.

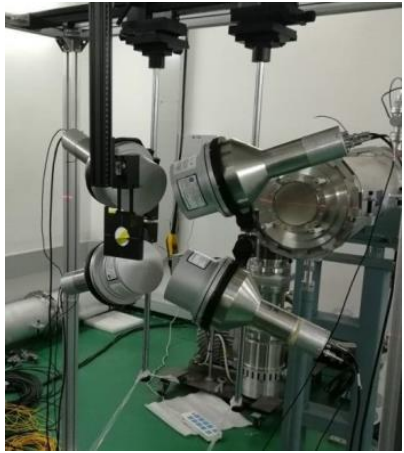

**Fig. 4.** C$_6$D$_6$ detectors online for experiment

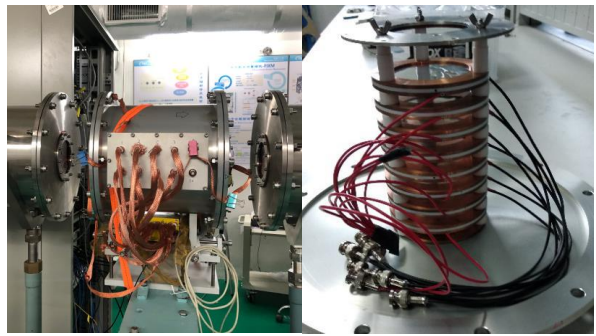

**Fig. 5.** FIXM spectrometer online and inside the device

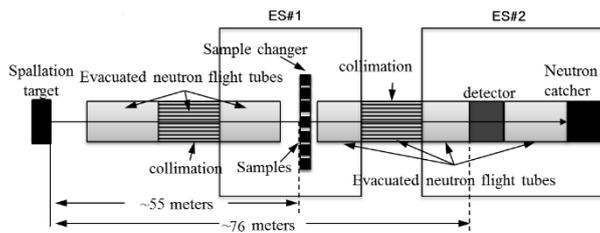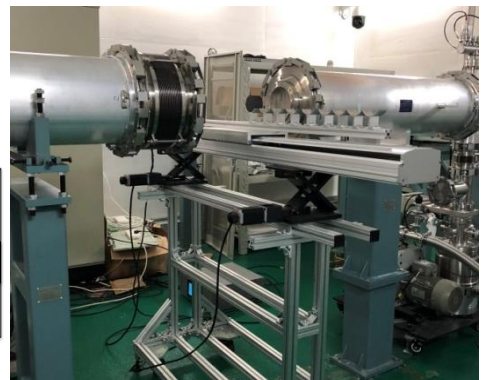

**Fig. 6.** Schematic for the NTOX spectrometer (upper) and sample changer (lower)

### 3.3 NTOX

The NTOX is depicted in Fig. 6. It is composed of two major parts: a sample auto-changing system located in ES#1 and a detector located in ES#2. The total cross-section measurements are conducted using the transmission measurement method. The distance of approximately 20 m between the sample and the detector and Collimator#2 (also the shielding) ensure that almost all the scattered neutrons from elastic reactions of the sample will be cleaned so as not to produce a background signal at the detector. Different types of detectors are planned for detecting neutrons in different energy ranges to assure high efficiency and high precision; however, the FIXM detector with multiple $^{235}$U and $^{238}$U samples was used in the early experiments. A dedicated multi-layer ionization chamber with $^{235}$U/$^{238}$U/$^{10}$B samples is ready for experiments, and $^{10}$B is helpful in determining the cross-sections in the low energy region with higher precision. Other types of detectors, such as scintillators, are still under development.

### 3.4 LPDA

The LPDA [14-15], as shown in Fig. 7, is a 16-unit detector array of ΔE-ΔE-E telescopes in a large vacuum chamber. Each unit has an MWPC (multi-wire proportional chamber)-type gas chamber for the low-energy ΔE measurement, a silicon detector for the higher-energy ΔE measurement, and a CsI(Tl) scintillator to measure the total energy. It measures the light charged particles generated in the (n, x) reactions. The relatively large solid coverage allows a high detection efficiency, and the telescope structure is essential for the wide detection energy range of the light charged particles, e.g., 0.5–100 MeV for protons. A complex gas-flow and pressure control system for the vacuum chamber is important, which allows the MWPC detectors to have very thin windows. The rotatable detector support plate and sample auto-changing system make the spectrometer flexible for different experiments. One can even put a lead block here to absorb gamma rays for other types of experiments at ES#2 or put the samples here for the total cross-section measurements fully in vacuum with low-energy neutrons. Although the whole system was completed in July 2020, the spectrometer has been employed since 2018 with simpler configurations, e.g., a 15-unit silicone array, multi-unit ΔE-E telescopes (MWPC+Si).

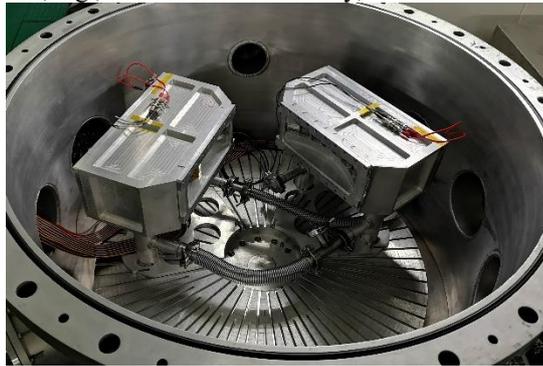

**Fig. 7.** LPDA spectrometer with the chamber cover open; 16 telescopes grouped in two boxes

*3.5 GTAF-II*

GTAF-II is the upgraded version of the GTAF [11] that was constructed around 2006 at the China Institute of Atomic Energy (CIAE), and it initially aimed for the experiments with neutrons produced by the HI-13 Tandem accelerator; GTAF-II was implemented at Back-n in 2019. It maintains the main structure with 40 units of $BaF_2$ detectors. The solid angle coverage is approximately 90% of 4π, considering that 2 of 42 spherical units were left out for the neutron beam pass. With the upgrading, a completely new readout electronics system (the common electronics will be discussed in Section 3.6) replaced the old one, which is more suitable for a high-intensity white neutron beam. A new mechanical support, including the detector skeleton shell and movable base support, was implemented to fit the Back-n end-station condition (see Fig. 8). An inner neutron absorber shell in polyethylene containing borium-10 will be added in the near future. The on-site installation at Back-n was completed in July 2019, and the tests with beams and the first experiments were subsequently conducted.

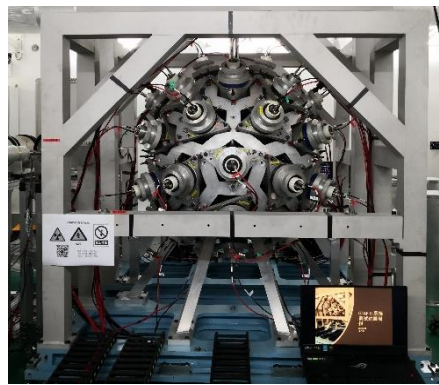

Fig. 8. GTAF-II during experiment at ES#2

*3.6 Generalized readout electronics and data acquisition*

The generalized readout electronics (or common electronics), which is adaptable to any of the above spectrometers when it is set up for online experiments, was designed based on the PXIe frame, and the wave-form digitization is based on the folding-ADC and FPGA techniques. It has a digital resolution of 12 bits and a sampling rate of 1 GSPS [31-32]. A full-digital trigger system that can be programmed according to different experiments or detectors can be used to reduce data volume to be transferred and recorded by the data acquisition system (DAQ). The system has been built up in steps, from 20 signal channels in the beginning to 64 channels in 2020, which can meet the requirements of all the above spectrometers and some user-owned detectors. The DAQ was designed to have a transferring baud rate and storing speed up to about 400 Mbps and 200 Mbps, respectively. The raw data is stored for a limited period locally and then transferred to the CSNS computation center for long-term storage through a 1 Gbps ethernet connection. The DAQ can manage the data acquisition of two or three parallel experiments, and it has expansibility to meet the requirements of larger data in the future.

*3.7 Gated CMOS camera for neutron imaging*

A gated CMOS camera system that was initially developed for beam spot measurements was frequently used for neutron imaging experiments (see Fig. 9). Its good space resolution and easy adjustment of the time-gating window enables the camera to provide contrast imaging for different neutron energies. When the time window is set at the characteristic resonances of a nuclide of the sample, it can reflect the information of the nuclide distribution in the sample, which is the so-called neutron resonant imaging method. The time windows for the resonances can be determined using the NTOX spectrometer or the total cross-section measurements.

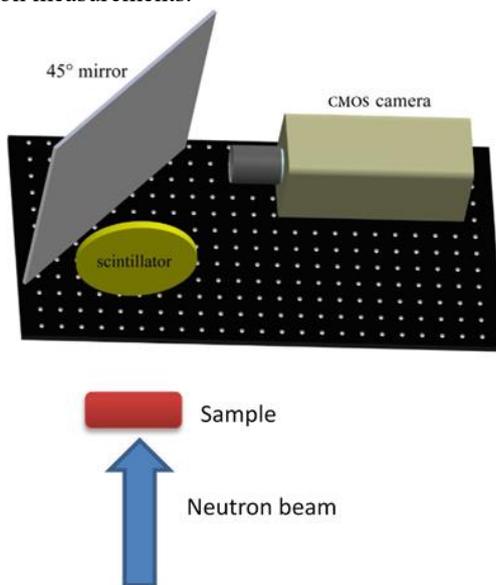

Fig. 9 Gated CMOS for neutron imaging

## 4. Applications

*4.1 Nuclear data measurements*

After the detailed measurements on the neutron beam characteristics and neutron/gamma background during the commissioning phase, different types of experiments at Back-n have commenced from April 2018, and most of the beam time has been for nuclear data measurements. With the five spectrometers that are currently available for measurements on (n, f) cross-sections, (n, tot) cross-sections, (n, $\gamma$) cross-sections, and (n, lcp) cross-sections and user-owned HPGe detectors for in-beam gamma spectrum or inelastic cross-section measurements, approximately 40 nuclides have been measured. These experiments demonstrate that the Back-n is an excellent TOF facility for nuclear data measurements. The measurements cover basic nuclear data, nuclear data for nuclear engineering, nuclear structure, and nuclear astrophysics.

*4.2 Irradiation tests*

A neutron beam of high flux and wide energy range, spanning from the order of meV to hundreds of MeV, can be used for different irradiation tests, including the single event effects of atmosphere neutrons in microchips and electronic boards, radiation damage of crystals and semiconductors, and material modification. There is a particularly large and increasing demand for chip irradiation tests in China that require high-flux neutrons above 1 MeV. At Back-n, the neutron beam at approximately 45% with energy higher than 1 MeV and a total flux of approximately $2\times10^7$ n/cm$^2$/s (ES#1) makes it among the best irradiation facilities in the world for chip irradiation tests. In many cases, the irradiation tests can be performed at ES#2 in the parasitic mode when nuclear data measurements are conducted at ES#1. The newly added irradiation facility just upstream of the neutron shutter can provide even higher neutron flux (approximately $8\times10^7$ n/cm$^2$/s) for irradiation tests in the fully-parasitic mode.

*4.3 Detector calibrations and tests*

The Back-n neutron beam has a very wide energy spectrum, high flux, and good time resolution; thus, it is very suitable for the calibration and test of neutron detectors. The dedicated single bunch operation mode has been employed for some of the tests. The two-dimensional double-bunch unfolding method, which is unfolding without knowing the detection efficiency with respect to neutron energy, is important for calibration in the MeV range and is still under development. New efforts in collaboration with the Radiation Metrology team at CIAE are underway to make the beamline a standard neutron calibration facility.

*4.4 Neutron imaging and element analysis*

Neutron imaging is a very important and widely used technique. Most neutron imaging facilities utilize thermal neutrons at reactor or spallation neutron sources, and much less use fast neutrons based on small or medium scale accelerators. At Back-n, one can perform imaging experiments with white neutrons covering the entire neutron resonant range for all elements, which is a novel technique. Because the resonant energies for different nuclides are characteristic, one can highlight the inner structures corresponding to different nuclides in a composite sample by showing the transmission images at the resonant energies together with the simple transmission imaging. We initially demonstrated the method using a gated CMOS camera. The combined neutron and gamma imaging method was also tested at Back-n [33]. For the neutron resonant imaging method, a more efficient detection system is under development to address the wide energy range.

With a white neutron beam, different element analysis methods that require significantly lower neutron flux than the resonant imaging method can be considered. At Back-n, we have successfully demonstrated the transmission method, which is very similar to total cross-section measurements and can identify nuclides in composite samples. Other methods, such as prompt gamma spectra measurements and neutron capture measurements are also being considered.

## 5. First years' operation

Since its opening to general users in September 2018, Back-n has been operating very efficiently, owing to the high reliability and increasing beam power of the accelerator complex. Approximately 50 experiments covering all the applications mentioned in Section 4 and beam time of more than 4000 hours per year have been allocated with a very low failure rate of less than 100 hours per year for the facility itself. These include users' experiments and machine studies, which are very important for us to master experimental methods with the powerful white neutron beam and the development of new spectrometers and methods. A user committee reviews the open proposals and the Back-n collaboration team can approve small projects of less than 48 hours and urgent experiments.

## 6. Conclusions

The Back-n white neutron source has been in operation since March 2018. As the first high-performance white neutron source in China, it provides a neutron beam suitable for TOF measurements and irradiation tests. With its high flux of up to $10^7$ n/cm$^2$/s at 55 m from the target, its wide energy range, from 0.5 eV to 200 MeV, and a good time resolution of a few per-mille, Back-n is among the leading white neutron sources in the world, especially its intensity is the highest. With its wide applications, the neutron source places emphasis on research in nuclear data measurements. Five of the eight planned spectrometers for nuclear data measurements have been put into operation. The operation experience in the last two years has demonstrated that the facility works with a high reliability and is able to conduct high-level experiments; however, more efforts to master the experimental methods and develop new methods are still in progress.

## Acknowledgements

This work is jointly supported by the National Key Research and Development Program of China (Project: 2016YFA0401600), National Natural Science Foundation of China (Projects: 11235012, 12035017), the CSNS Engineering Project, and the Back-n Collaboration Consortium fund. The authors would like to thank their colleagues from CSNS and the five supporting institutions for all kinds of support.